\documentclass[runningheads]{llncs}
\pdfoutput=1
\usepackage{graphicx}
\usepackage{hyperref}

\newcommand{\repeatthanks}{\textsuperscript{\thefootnote}}

\begin{document}

\title{When Weak Becomes Strong: \\
Robust Quantification of White Matter Hyperintensities in Brain MRI scans}
\titlerunning{When Weak Becomes Strong}

\author{Oliver Werner\inst{1} \thanks{Corresponding author. E-mail address: \email{o.werner@erasmusmc.nl}.}
\and Kimberlin M.H. van Wijnen \inst{1}
\and Wiro J. Niessen \inst{1,2, 4}
\and Marius de Groot \inst{1, 4}
\and Meike W. Vernooij \inst{1,3}
\and Florian Dubost \inst{1}\thanks{Equal contribution.} 
\and Marleen de Bruijne \inst{1,5}\repeatthanks}

\authorrunning{O. Werner et al.}

\institute{
BIGR, Department of Radiology and Nuclear Medicine, Erasmus MC, Rotterdam, The Netherlands
\and Department of Imaging Science and Technology, Faculty of Applied Sciences, Delft University of Technology, Delft, The Netherlands
\and Department of Epidemiology, Erasmus MC, Rotterdam, The Netherlands
\and Department of Medical Informatics, Erasmus MC , Rotterdam, The Netherlands
\and Department of Computer Science, University of Copenhagen, Denmark}

\maketitle      

\begin{abstract}

    To measure the volume of specific image structures, a typical approach is to first segment those structures using a neural network trained on voxel-wise (strong) labels and subsequently compute the volume from the segmentation. A more straightforward approach would be to predict the volume directly using a neural network based regression approach, trained on image-level (weak) labels indicating volume.
    
    In this article, we compared networks optimized with weak and strong labels, and study their ability to generalize to other datasets. We experimented with white matter hyperintensity (WMH) volume prediction in brain MRI scans. Neural networks were trained on a large local dataset and their performance was evaluated on four independent public datasets. We showed that networks optimized using only weak labels reflecting WMH volume generalized better for WMH volume prediction than networks optimized with voxel-wise segmentations of WMH. 
    The attention maps of networks trained with weak labels did not seem to delineate WMHs, but highlighted instead areas with smooth contours around or near WMHs.
    By correcting for possible confounders we showed that networks trained on weak labels may have learnt other meaningful features that are more suited to generalization to unseen data. 
    Our results suggest that for imaging biomarkers that can be derived from segmentations, training networks to predict the biomarker directly may provide more robust results than solving an intermediate segmentation step.

\end{abstract}
\section{Introduction}
Quantitative imaging biomarkers are important indicators and predictors of diseases. 
Manually annotating images to measure quantitative biomarkers is often costly, time-consuming, and error-prone. 
Convolutional neural networks (CNNs) are often used to automate the extraction of quantitative biomarkers.
Generally, a CNN is trained with voxel-wise (strong) labels to perform a segmentation task and subsequently quantitative biomarkers such as e.g. volume or shape are derived \cite{guerrero2018wmh,moeskops2018brain}. A more straightforward approach would be to optimize the method to predict the quantitative biomarker directly. 
Weak, image-level labels are often used to either leverage more training data or to provide additional regularization \cite{Dubost2017a,Feng2017,Jia2017,Kervadec2019a,kervadec2019constrained}.
Jia et al. \cite{Jia2017} proposed a weakly supervised method  for segmentation of cancerous regions in histopathology images, optimized to predict image-level binary labels and using a rough estimation of the cancerous proportion as area constraint in the loss function.
Dubost et al. \cite{Dubost2017a} proposed a weakly supervised approach for brain lesion detection optimized using weak labels. A U-Net-like architecture with  a global pooling step was used during training to allow for optimization on the number of brain lesions per scan and during inference the network without global pooling layer was used to predict the location of brain lesions.
Kervadec et al. \cite{Kervadec2019a} proposed a semi-supervised approach for segmentation in which weak labels were used as a constraint in the loss function and as a way to train on unlabeled data. First a regression network was trained to predict the total volume (computed from the segmentations). This trained network was then used as a constraint measure for a fully supervised network that learns to segment with both labeled data (segmentations) and unlabeled data (predicted total volume).

In these methods weak labels were used as either surrogate or additional objective for segmentation or detection. However in some applications, segmentation is not necessary and the weak label could then become the primary objective. White matter hyperintensity (WMH) volume quantification is a clear example of such an application. While in most clinical studies only WMH volume is used for the analyses, most automated methods are optimized to segment WMHs \cite{ghafoorian2017wmh,guerrero2018wmh,kuijf2019standardized,moeskops2018brain}. 
WMHs are areas on fluid-attenuated inversion recovery (FLAIR) and T2 weighted MRI scans that are hyperintense in the white matter. There are various etiologies that can lead to WMHs, e.g. WMHs of vascular nature (referred to as WMHs of presumed vascular origin) and of inflammatory nature such as multiple sclerosis lesions (MS lesions) \cite{cSVDmarkers,strive}. Depending on the dataset (clinical vs population study e.g.) the number, volume and etiology of WMHs can vary. 

In this study we investigated the advantages of weak versus strong labels for white matter hyperintensity quantification in brain MRI. We optimized our networks on a large local dataset and evaluated the performance on a separate part of this local dataset and on four public datasets with different acquisition parameters. We showed that networks trained with only weak labels could provide both better performance of  WMH volume prediction and better generalization to other datasets than networks trained with only strong labels. We visualised the attention maps to compare the features learned between the different label types. Furthermore, we showed that the volume prediction performance was to a large extent independent of possible confounders such as the ventricle volume. 

\section{Methods and Materials}
In this section, we describe the datasets, the networks trained with weak labels, the networks trained with strong labels, training details, and evaluation.

\subsection{Datasets} \label{sec:Data}
We used data from the Rotterdam Scan Study (RSS) \cite{ikram2017rotterdam} to train the networks, and used a separate part of the RSS dataset combined with four smaller public datasets from WMH and MS lesion segmentation challenges to evaluate the networks.

The RSS dataset is a large population-based imaging study and the cohort used contained 4334 FLAIR scans. All scans were acquired on a 1.5 T GE MRI scanner with a reconstructed voxel resolution of $0.49 \times 0.49 \times 2.5 \mbox{mm}^{3}$. Further information on the acquisition of these scans is discussed by De Leeuw et al. \cite{de2001prevalence}. WMHs were annotated in all scans using an automated method using a FLAIR-based threshold to classify certain voxels as WMH \cite{de2009white}. All segmentations were subsequently inspected and manually corrected by a pool of experienced raters.

Furthermore, we used 60 FLAIR scans of the WMH Segmentation (WmhSeg) challenge \cite{kuijf2019standardized}, 21 FLAIR scans of the Longitudinal Multiple Sclerosis Lesion Segmentation (LongMSLes) challenge \cite{carass2017longitudinal}, 15 FLAIR scans of the MS segmentation (MSSEG) challenge \cite{commowick2018objective} and 15 FLAIR scans of the MS lesion segmentation (MSLes) challenge \cite{styner20083d}. All scans were acquired from different subjects, except for the LongMSLes challenge dataset, which contained 4 or 5 scans per subject. Apart from several scans from the MSSEG challenge which were acquired on a 1.5 tesla MRI scanner from Siemens, all scans were acquired with 3 tesla MRI scanners from mainly Philips and Siemens and one MRI scanner from GE. Voxel resolutions ranged from 0.5 to 1.03 mm in axial plane and from 0.7 to 3 mm in slice thickness. For the WmhSeg challenge only WMHs of presumed vascular origin were annotated, for the other challenges only MS lesions were annotated. For every challenge, one or two expert observers or radiologists segmented all WMHs, except for the MSSEG challenge which provided a consensus segmentation derived from segmentations of seven radiologists. 

All challenge scans were preprocessed to resemble the scans from the local dataset in terms of voxel spacing and cropping. Furthermore, brain extraction and bias correction were applied using N4\cite{tustison2010n4itk}. For every scan the total volume of WMHs was obtained by counting the number of voxels included in the WMH segmentations in the full scan.

\begin{figure}[b!]
    \centering
    \includegraphics[width=1.0\textwidth]{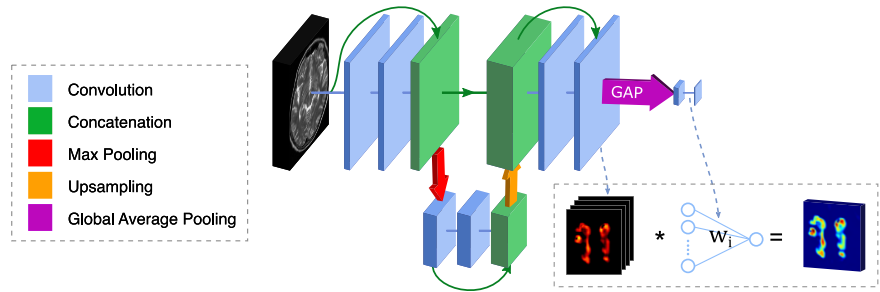}
    \caption{\textbf{Overview of the network optimized with weak labels.} A shallow U-Net like structure was used with a global pooling (GP) layer at the end. After every two convolutions, the feature maps before the convolutions are concatenated. The weights learned by the dense layer after the GP layer are used to weigh the last feature maps to create an attention map.}
    \label{Experiments:architecture_gpunet}
\end{figure}

\subsection{Network Architecture}
The models using weak and strong labels differ slightly as they were optimized with different label types. Both networks trained with strong and weak labels were kept as similar as possible for optimal comparison. The two architectures are discussed below. 

\subsubsection{Regression Network Trained with Weak Labels.}

 This network architecture was inspired by GP-Unet \cite{Dubost2017a} and was trained using image-level labels, namely total WMH volume per scan. GP-Unet combines a U-Net architecture generally used for segmentation \cite{ronneberger2015u} with the global pooling (GP) method proposed in \cite{zhou2016learning}. This architecture improves  precision thanks to the upsampling layers and skip connections of the U-Net. The global pooling layer enables supervision with image-level labels. The loss used during training was mean squared error. During inference, the network outputs the predicted total WMH volume. Furthermore, the weights learned after the global pooling layer are multiplied with the decoded feature maps before global pooling to create an attention map of the network, as shown in Figure \ref{Experiments:architecture_gpunet}. For every image, a WMH segmentation was obtained by thresholding this attention map with the threshold that provides a sum of foreground voxels equal to the predicted WMH volume in that image. After each convolutional layer except for the last layer, a ReLu activation \cite{nair2010rectified} was used.

\subsubsection{Segmentation Network Trained with Strong Labels.}
The structure of the voxel-wise optimized network was similar to the aforementioned GP-Unet, excluding the global pooling layer at the end, leading to a shallow U-Net architecture. A sigmoid activation was added at the end. During training, this network was optimized using the Dice coefficient. It was trained using voxel-wise labels containing the WMH segmentations. The output of the network was thresholded at 0.5 to achieve the predicted segmentations. The total WMH volume was computed by summing over the number of foreground voxels of the predicted segmentation mask.

\subsection{Training Details}
The local data was randomly split in a train (60\%), a validation (20\%), and a test (20\%) set containing 2602, 866 and 866 images, respectively. Each experiment was repeated three times, where the only difference was the random initialization of the network weights. Adadelta was used for optimization \cite{Zeiler2012a}. Batches of four full brain images of $112 \times 128 \times 32$ voxels were used as input. The images were percentile normalized (1\% and 99\%) between 0 and 1.  Standard on-the-fly data augmentation was applied during training, consisting of translation (uniform sampled between $\pm$0.2 of the dimension length), rotation ($\pm$54 degrees), and flipping. All code was written in Python, and Keras was used with Tensorflow as backend.

\subsection{Evaluation}
The networks are evaluated on their performance in WMH volume prediction and WMH segmentation. For all networks the predicted total WMH volume was compared to the WMH volume computed from the annotations using intraclass correlation (ICC), in this case ICC(2,1) \cite{koo2016guideline}. Furthermore, for all networks, the predicted segmentations are compared to the annotated WMH segmentations using Dice score. 

Additionally, for the networks trained with weak labels we investigated to what extent the volume prediction could be explained by measuring other factors that may have strong correlation with WMH volume, including ventricle volume, white matter volume, and cerebrospinal fluid  (CSF) (outside the ventricles). 
To this end, we modeled the relationship between predicted total WMH volume and total volume computed from the annotations using linear regression with ordinary least squares and adjusted for these confounders. The coefficient of determination was computed with and without these confounders and the significance was assessed by the p-value. The confounder volumes were extracted from the brain MRI scans using FreeSurfer \cite{desikan2006automated}.

\section{Results}

All models were evaluated on the test set of the local dataset, containing 866 scans. While models optimized with weak labels performed poorly in segmentation, with an average Dice score of 0.08 compared to the 0.78 of the network optimized with strong labels, they performed very well in volume prediction. The average ICC between the predicted WMH volume and the annotated WMH volume for the models trained using weak labels was 0.99, while for the strong label models this was substantially lower with a value of 0.81. 

\begin{figure}[ht!]
    \centering
    \includegraphics[width=0.8\textwidth]{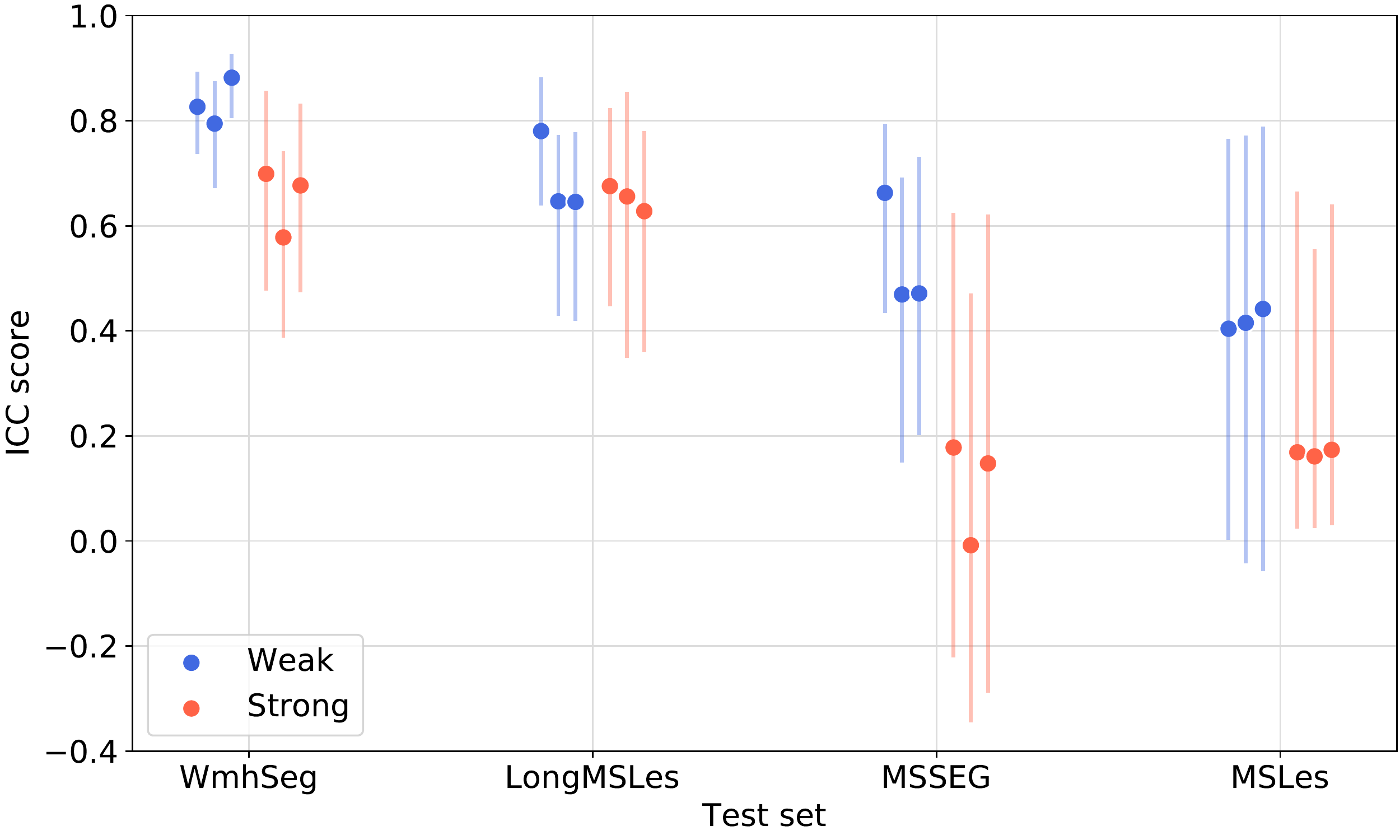}
    \caption{\textbf{Volume prediction results on the public datasets.} The y-axis shows the average intraclass correlation coefficient(ICC) between the predicted and ground truth WMH volume. Each point corresponds to a network trained with a different random initialization of the weights. The lines indicate the corresponding confidence intervals (5\% and 95\%). The datasets are displayed on the x-axis, details can be found in section \ref{sec:Data}.}
    \label{Experiment:generalization_scatter}
\end{figure}

Figure \ref{Experiment:generalization_scatter} shows the average ICC between total annotated WMH volume and predicted WMH volume by the models trained on the local data and applied to the four public datasets. 
The performance of models trained with weak labels on total WMH volume prediction was consistently similar or higher than the performance of models trained with strong labels for all experiments.

\begin{figure}[hb!]
    \centering
    \includegraphics[width=0.75\textwidth]{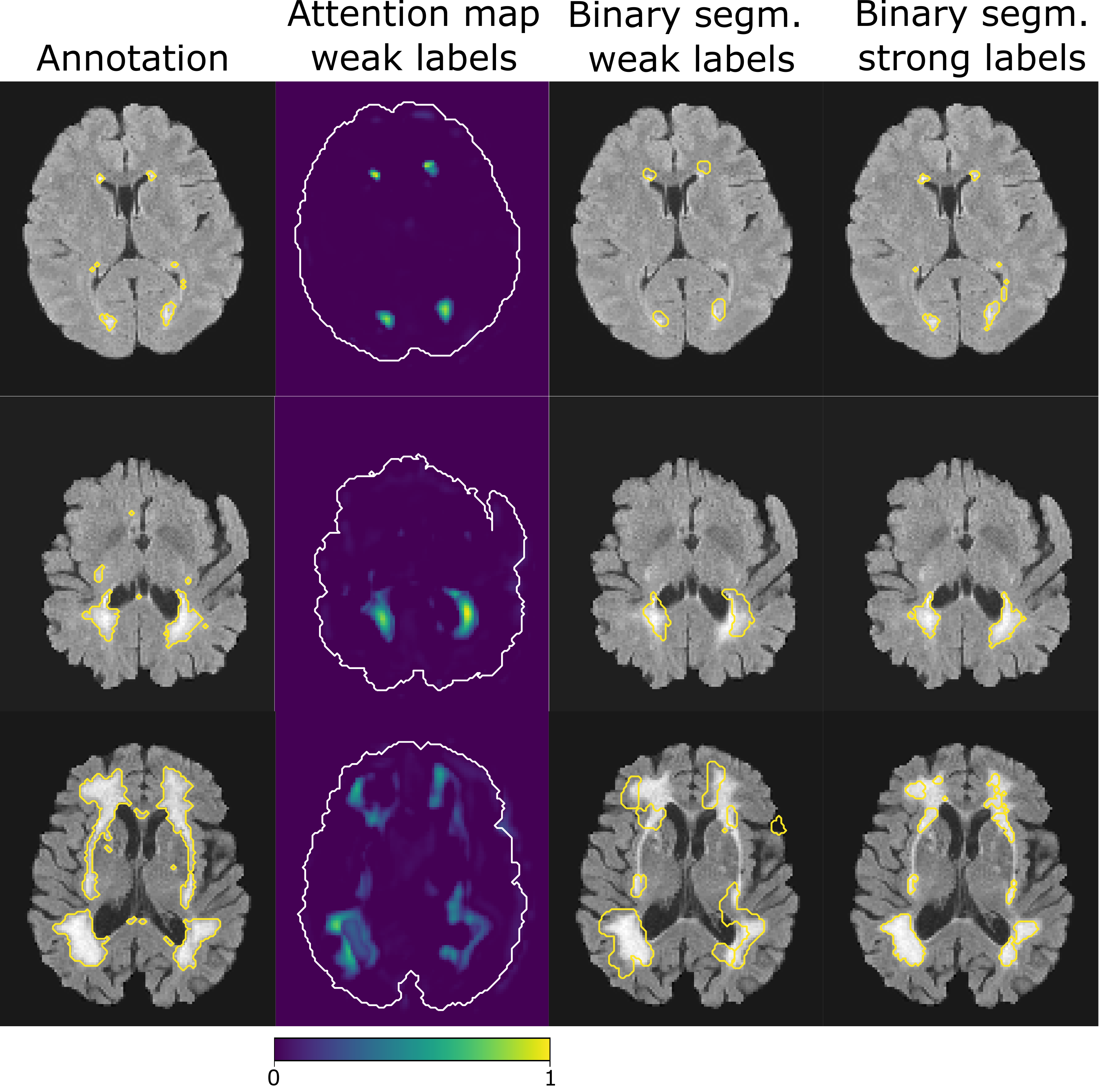}
    \caption{\textbf{Examples of predictions on the test set of the local dataset}. The rows show different subjects. The columns from left to right show an axial slice of the MRI scan overlaid with the annotated WMH contours, the attention map of a network trained with weak labels (contour of brain shown), the corresponding predicted segmentations overlaid on the MRI scan and the last column shows the predicted segmentations by a network trained with strong labels overlaid on the MRI scan. The outputs of the networks trained with strong labels were almost binary due to the sigmoid activation, and were almost identical to the binary segmentations.}

    \label{Experiments:feature_maps}
\end{figure}

Figure \ref{Experiments:feature_maps} shows the thresholded attention maps of networks trained with weak labels. The predicted WMH segmentation of the model trained with strong labels was very similar to the annotated WMH segmentation for the large WMH regions. 
The attention map generated by the network trained with weak labels seemed to focus on WMHs. The areas of attention seemed larger, more blurry and smoother than the annotated segmentation, and sometimes seemed shifted.

The least square multiple regression between the annotated WMH volumes and the WMH volumes predicted by models trained with weak labels showed a strong, significant correlation (p $<$ 0.01), a Pearson correlation of 0.99 and coefficient of determination of 0.98. The Pearson correlation between ground truth WMH volume and the possible confounders was lower, with a correlation of 0.41 for the ventricles, 0.47 for the white matter, and 0.45 for the CSF.
After adjusting for the volume of ventricles, white matter, and CSF, the correlation between ground truth and predicted WMH volumes was still significant (p $<$ 0.01) and coefficient of determination stayed at 0.98. 

\section{Discussion and Conclusion}
Networks optimized with weak labels predicted the WMH volume more accurately than networks optimized with strong labels, both in the local training dataset and in the external datasets. The networks optimized with weak labels could not be used to segment WMHs, but did seem to focus on areas near the WMHs. 
The correlation between WMH ground truth volumes and WMH volume prediction of the networks optimized with weak labels remained strong and significant after correcting for possible confounders such a ventricle volume, white matter volume and CSF volume. These results suggest that networks optimized with weak labels may be able to identify imaging features that are more suited to generalization to other datasets. 

Networks trained with weak labels predicted WMH volume more accurately than the networks trained with strong labels. 
An explanation for this could be that networks trained with weak labels were optimized on the target quantity, namely the WMH volume, while networks trained with strong labels were optimized for an intermediary task, the WMH segmentation. The correlation between the WMH segmentation performance in Dice overlap and the WMH volume prediction performance in ICC was not excellent for networks trained with strong labels (Pearson coefficient of 0.71, Spearman 0.87). 
Networks that predicted the segmentation the most accurately were not the models that also predicted the volume most accurately. This underlines the differences between the two tasks and questions the validity of optimizing on WMH segmentations if the objective is to quantify WMH volume.
Another explanation could be that networks optimized with strong labels are forced to predict a binary label for every voxel. The labels of some of these voxels, e.g. at the border of the WMH, are uncertain, and forcing the network to classify them may enforce the learning of systematic biases present in the annotations and impede the generalizability. For example, networks might rely on the intensity threshold used by semi-automated WMH annotations methods to create the networks' training ground truth segmentations. Networks optimized with weak labels are not forced to predict voxel-level labels and may be more suited to processing uncertain voxels.

The attention maps of networks trained with weak labels did not seem to delineate WMHs, but highlighted instead areas with smooth contours around or near WMHs. Furthermore, the statistical test showed that the features were not derived from obvious confounders. Lastly, the volume prediction of these models was better in all test datasets. These observations could indicate that the features learned using weak labels are more robust and generalize better than the features learned using strong labels. 

A drawback of optimizing networks with weak labels is that the interpretation of features learnt by networks is more difficult. We corrected for the most natural confounders, but less obvious confounders may have been omitted in our analysis and may have influenced the network predictions. Applied to another imaging quantification problem, researchers have to actively consider how potential confounders can influence the predictions of the networks. 

As we showed that both label types have different advantages, it could be beneficial to optimize on both label types simultaneously. Weak and strong labels have been used jointly with promising results in weakly and semi-supervised methods \cite{Jia2017,Kervadec2019a}. 

The difference of WMH volume prediction performance between networks trained with weak labels and network trained with strong labels seemed to be higher in datasets that are most dissimilar to the training dataset. This dissimilarity could be due to e.g. voxel spacing, difference in etiology of the WMHs (e.g. vascular or inflammatory WMHs), differences in patient population or annotation style.
WMHs of vascular origin were the most common type of WMHs in the local dataset. This might be the reason that from all external datasets the best results were obtained in WMHSeg, as this dataset contained WMH of vascular origin while the other three datasets contained MS lesions. 

As the volume was still extracted from voxel-wise annotations, there was no benefit in terms of cost and effort. Severity scores, like the Fazekas score \cite{fazekas1987mr}, are more time-efficient to annotate and can also be used for training. A pilot experiment with severity scores was conducted, in which the severity score was derived from categorizing the WMH volume from the local dataset into 5 classes based on the value. The model achieved an ICC of 0.92 when evaluated on the test set, which shows there is potential for using severity scores as a weak label. 

Our results suggest that when the objective is to quantify total WMH volume, training a network to predict this total WMH volume might be more suitable than training a segmentation network and deriving the total WMH volume from the resulting segmentation.
More generally, for imaging biomarkers that can be derived from segmentations, training networks to predict the biomarker directly may provide more robust results than solving an intermediate segmentation step.

\subsubsection{Acknowledgments}
This research was funded by the Netherlands Organisation for Health Research and Development (ZonMw), project 104003005, as well as by the Dutch Technology Foundation STW and Quantib, project number P15-26.

\bibliographystyle{splncs04}
\bibliography{main}

\end{document}